\documentclass[twocolumn]{aastex61}

\pdfoutput=1 
\usepackage{amsmath,amstext}
\usepackage[T1]{fontenc}
\usepackage{apjfonts} 
\usepackage[figure,figure*]{hypcap}

\usepackage{CJKutf8}
\AtBeginDvi{\input{zhwinfonts}} 
\newcommand{\Chi}[2]{%
  \csname CJK*\endcsname{UTF8}{zhsong}%
    \CJKchar{#1}{#2}%
  \csname endCJK*\endcsname
}
\DeclareUnicodeCharacter{9673}{\Chi{"96}{"73}}
\DeclareUnicodeCharacter{82F1}{\Chi{"82}{"F1}}
\DeclareUnicodeCharacter{540C}{\Chi{"54}{"0C}}
\DeclareUnicodeCharacter{6797}{\Chi{"67}{"97}}
\DeclareUnicodeCharacter{7701}{\Chi{"77}{"01}}
\DeclareUnicodeCharacter{6587}{\Chi{"65}{"87}}

\newcommand{\gsim}{\lower.7ex\hbox{$\;\stackrel{\textstyle>}{\sim}\;$}}
\newcommand{\lsim}{\lower.7ex\hbox{$\;\stackrel{\textstyle<}{\sim}\;$}}

\newcommand{\PS}{PS1}
\newcommand{\OSS}{Outer Solar System}

\newcommand{\NAME}{2010~JO$_{179}$}

\shorttitle{A Dwarf Planet in the 21:5 Resonance}
\shortauthors{Holman et al.}

\submitjournal{AJ, September 15 2017}

\begin{document}

\title{A Dwarf Planet Class Object in the 21:5 Resonance with Neptune}


\author[0000-0002-1139-4880]{Matthew~J.~Holman}
\affiliation{Harvard-Smithsonian Center for Astrophysics, 60 Garden St., MS 51, Cambridge, MA 02138, USA}
\correspondingauthor{Matthew~J.~Holman}
\email{mholman@cfa.harvard.edu}

\author[0000-0001-5133-6303]{Matthew~J.~Payne} 
\affiliation{Harvard-Smithsonian Center for Astrophysics, 60 Garden St., MS 51, Cambridge, MA 02138, USA}

\author[0000-0001-6680-6558]{Wesley~Fraser} 
\affiliation{Astrophysics Research Centre, School of Mathematics and Physics, Queen’s University Belfast, Belfast BT7 1NN, UK}

\author[0000-0002-1708-4656]{Pedro~Lacerda} 
\affiliation{Astrophysics Research Centre, School of Mathematics and Physics, Queen’s University Belfast, Belfast BT7 1NN, UK}

\author[0000-0003-3257-4490]{Michele~T.~Bannister}
\affiliation{Astrophysics Research Centre, School of Mathematics and Physics, Queen’s University Belfast, Belfast BT7 1NN, UK}

\author{Michael Lackner} 
\affiliation{Institute for Applied Computational Science, Harvard University, Cambridge, MA, USA}

\author[0000-0001-7244-6069]{Ying-Tung~Chen }
\affiliation{Institute of Astronomy and Astrophysics, Academia Sinica; 11F of AS/NTU Astronomy-Mathematics Building, Nr. 1 Roosevelt Rd., Sec. 4, Taipei 10617, Taiwan}

\author[0000-0001-7737-6784]{Hsing~Wen~Lin }
\affiliation{Institute of Astronomy, National Central University, Taoyuan 32001, Taiwan}
\affiliation{Department of Physics, University of Michigan, Ann Arbor, MI 48109, USA}

\author{Kenneth~W.~Smith} 
\affiliation{Astrophysics Research Centre, School of Mathematics and Physics, Queen’s University Belfast, Belfast BT7 1NN, UK}

\author{Rosita~Kokotanekova} 
\affiliation{Max Planck Institute for Solar System Research, Justus-von-Liebig-Weg 3, 37077 Göttingen, Germany}

\author{David~Young} 
\affiliation{Astrophysics Research Centre, School of Mathematics and Physics, Queen’s University Belfast, Belfast BT7 1NN, UK}

\author{K. Chambers} 
\affiliation{Institute for Astronomy, University of Hawaii, 2680 Woodlawn Drive, Honolulu HI 96822, USA}

\author{S. Chastel} 
\affiliation{Institute for Astronomy, University of Hawaii, 2680 Woodlawn Drive, Honolulu HI 96822, USA}

\author{L. Denneau} 
\affiliation{Institute for Astronomy, University of Hawaii, 2680 Woodlawn Drive, Honolulu HI 96822, USA}

\author{A. Fitzsimmons} 
\affiliation{Astrophysics Research Centre, School of Mathematics and Physics, Queen’s University Belfast, Belfast BT7 1NN, UK}

\author{H. Flewelling} 
\affiliation{Institute for Astronomy, University of Hawaii, 2680 Woodlawn Drive, Honolulu HI 96822, USA}

\author{Tommy~Grav} 
\affiliation{Planetary Science Institute, Tucson, AZ 85719, USA}

\author{M. Huber} 
\affiliation{Institute for Astronomy, University of Hawaii, 2680 Woodlawn Drive, Honolulu HI 96822, USA}

\author{Nick~Induni} 
\affiliation{Harvard-Smithsonian Center for Astrophysics, 60 Garden St., MS 51, Cambridge, MA 02138, USA}

\author{Rolf-Peter Kudritzki} 
\affiliation{Institute for Astronomy, University of Hawaii, 2680 Woodlawn Drive, Honolulu HI 96822, USA}

\author{Alex~Krolewski} 
\affiliation{Harvard-Smithsonian Center for Astrophysics, 60 Garden St., MS 51, Cambridge, MA 02138, USA}

\author{R. Jedicke} 
\affiliation{Institute for Astronomy, University of Hawaii, 2680 Woodlawn Drive, Honolulu HI 96822, USA}

\author{N. Kaiser} 
\affiliation{Institute for Astronomy, University of Hawaii, 2680 Woodlawn Drive, Honolulu HI 96822, USA}

\author{E. Lilly} 
\affiliation{Institute for Astronomy, University of Hawaii, 2680 Woodlawn Drive, Honolulu HI 96822, USA}

\author{E. Magnier} 
\affiliation{Institute for Astronomy, University of Hawaii, 2680 Woodlawn Drive, Honolulu HI 96822, USA}

\author{Zachary~Mark} 
\affiliation{Harvard-Smithsonian Center for Astrophysics, 60 Garden St., MS 51, Cambridge, MA 02138, USA}

\author{K.~J. Meech} 
\affiliation{Institute for Astronomy, University of Hawaii, 2680 Woodlawn Drive, Honolulu HI 96822, USA}

\author{M. Micheli} 
\affiliation{Institute for Astronomy, University of Hawaii, 2680 Woodlawn Drive, Honolulu HI 96822, USA}

\author{Daniel~Murray} 
\affiliation{Department of Physics, University of Wisconsin, Milwaukee, WI, 53211, USA}

\author{Alex Parker} 
\affiliation{Southwest Research Institute, Department of Space Studies, Boulder, CO, 80302, USA}

\author{Pavlos~Protopapas} 
\affiliation{Institute for Applied Computational Science, Harvard University, Cambridge, MA, USA}

\author{Darin~Ragozzine} 
\affiliation{Brigham Young University, Department of Physics and Astronomy, Provo, UT 84602, USA}

\author{Peter~Veres} 
\affiliation{Harvard-Smithsonian Center for Astrophysics, 60 Garden St., MS 51, Cambridge, MA 02138, USA}

\author{R. Wainscoat} 
\affiliation{Institute for Astronomy, University of Hawaii, 2680 Woodlawn Drive, Honolulu HI 96822, USA}

\author{C. Waters} 
\affiliation{Institute for Astronomy, University of Hawaii, 2680 Woodlawn Drive, Honolulu HI 96822, USA}

\author{R. Weryk} 
\affiliation{Institute for Astronomy, University of Hawaii, 2680 Woodlawn Drive, Honolulu HI 96822, USA}

%
%
%
%
%
%
%
%
%


\begin{abstract}
We report the discovery of a $H_r = 3.4\pm0.1$ dwarf planet candidate by the Pan-STARRS Outer Solar System Survey.
\NAME\ is red with $(g-r)=0.88 \pm 0.21$, roughly round, and slowly rotating, with a period of $30.6$ hr. Estimates of its albedo imply a diameter of 600--900~km. 
Observations sampling the span between 2005--2016 provide an exceptionally well-determined orbit for \NAME, with a semi-major axis of $78.307\pm0.009$ au; distant orbits known to this precision are rare.
We find that \NAME\ librates securely within the 21:5 mean-motion resonance with Neptune on hundred-megayear time scales, joining the small but growing set of known distant dwarf planets on metastable resonant orbits. These imply a substantial trans-Neptunian population that shifts between stability in high-order resonances, the detached population, and the eroding population of the scattering disk.
\end{abstract}

\section{Introduction}
\label{SECN:INTRO}
Dwarf planets in the trans-Neptunian region are remnant planetesimals from the protoplanetary disk of the Solar System. They constrain the large-diameter end of the trans-Neptunian object (TNO) size distribution, which is inferred from the observed luminosity function \citep{Petit:2008th,Brown:2008tp, Schwamb:2013cu,Fraser.2014}.
However, few are yet known:\footnote{\url{http://www.minorplanetcenter.net/db\_search}:
$a\geq30$, $q\geq30$ and $H_V\leq4$, as of 15 August 2017} $33$ with absolute magnitude $H_V~<~4$. 

TNOs are faint due to their $>30$ au heliocentric distances, requiring discovery surveys by $>1$ m aperture wide-field optical telescopes. 
Past wide-area surveys have completed the inventory of bright TNOs to $m_r \sim 19.5$ outside the galactic plane \citep{Tombaugh:1961p3095,Kowal:1989bd,Sheppard.2000,Trujillo.2003,Moody:2004us,Brown:2008tp,Brown.2015}. Substantial areas of sky have been surveyed to $m_r \sim 21.5$ and deeper \citep{Larsen:2001,Larsen:2007,Elliot:2005ju,Schwamb.2010,Sheppard.2011,Petit.2011,Rabinowitz.2012,Sheppard:2016jf,Petit:2017ju,Gerdes2017}. 

As often the brightest and thus easiest to detect of the worlds in the trans-Neptunian region, dwarf planets also provide a useful broad-brush indication of the phase space of their dynamical populations. They are key to exploring the fainter, large-semi-major-axis populations where the smaller TNOs are too faint to detect; for example, the 2003 discovery of (90377) Sedna indicated a substantial population of TNOs with large perihelion distances \citep{B04}. The $a \gtrsim 50$~au populations are all defined by their degree of gravitational interaction with Neptune: they include orbits librating in high-order mean-motion resonance; the scattering disk, on orbits actively interacting with Neptune; and the ``detached" TNOs, with perihelia $q \gtrsim 37$~au \citep{Gladman.2008}.
Recent discoveries include the 9:2 mean motion resonant object 2015 RR$_{245}$ with $H_r = 3.6$ and $a = 81.86 \pm 0.05$ au \citep{Bannister16}, the scattering disk TNO 2013 FY$_{27}$ with $H_V = 2.9$ and $a = 59$ au \citep{Sheppard:2016jf}, and the detached TNO 2014 UZ$_{224}$ with $H_V = 3.5$ and $a = 109 \pm 7$ au \citep{Gerdes2017}. 

The Panoramic Survey Telescope and Rapid Response System 1 Survey (Pan-STARRS 1, hereafter \PS) is well suited to the discovery of TNOs. \PS\ is a 1.8 m telescope on Haleakela in Hawai'i, with a dedicated 0\arcsec.258 pixel, 7 deg$^2$ optical imager \citep{Kaiser.2010,Chambers.2016}. 
The \PS~$3\pi$ survey \citep{Magnier.2013,Chambers.2016} repeatedly observed the sky north of Decl. $-30\arcdeg$ using a Sloan-like filter system \citep{Tonry.2012}, reaching typical single-exposure $5\sigma$ depths of $g_{\mathrm{P1}}=22.0$, $r_{\mathrm{P1}}=21.8$, and $i_{\mathrm{P1}}=21.5$ \citep[Table 11]{Chambers.2016}. 
\PS\ also observed within $\pm20\arcdeg$ of the ecliptic using the wide $w_{P1}$ filter ($m_{5\sigma} \sim 22.5$). 
The observing cadence of \PS\ is optimized for detection of inner Solar System minor planets, using the PS Moving Object Processing System (MOPS) \citep{Denneau.2013}. However, \PS's many visits permit detection of slower-moving ($\sim 3\arcsec/hr$) TNOs in the accumulated data.
\citet{W16} reported several hundred centaurs and TNOs from a search of the 2010 Feb 24 to 2015 July 31 \PS\ observations, which linked together detections within 60-day intervals. 

The PS1 \OSS\ (OSS) key project uses a novel linking solution, based on transforming topocentric observations to a heliocentric coordinate frame using an \emph{assumed} heliocentric distance.
Our initial search of the 2010-2014 \PS\ observations resulted in hundreds of candidates, $\sim 50\%$ of which were newly discovered TNOs \citep{Holman.2015}. 
These include unusual objects such as the highly-inclined centaur (471325) 2011 KT$_{19}$ \citep{Chen16}, as well as numerous new Neptune trojans \citep{Lin16}. 

Here we report the discovery of a $m_r \sim 21$ dwarf planet candidate at a barycentric distance of 55 au: \NAME.
We present the technique used to detect \NAME\ (\S~\ref{SECN:DET}), our observations (\S~\ref{sec:obs}), \NAME's physical properties (\S~\ref{SECN:PHYS}), the dynamical classification of its orbit (\S~\ref{SECN:DYN}), and the broader implications of its existence for our understanding of the distant, dynamically excited TNO populations (\S~\ref{SECN:CON}).

\section{TNO Discovery Technique}
\label{SECN:DET}

We developed our discovery pipeline to cope with the temporal sparsity of the Pan-STARRS data (see \citealt{Brown.2015} for an independent, alternative approach to detecting TNOs in sparse data sets).
Our pipeline operates on the catalogs of source detections found in the direct, undifferenced \PS\ exposures spanning 2010 to mid-2014 by the Image Processing Pipeline (IPP)~\citep{PS1_IPP, PS1_photometry}.  We eliminate any detection that coincides within $1\arcsec$ of a known stationary source. (A catalog of stationary sources was developed from the individual detections in the PS1 exposures; detections that occur near the same location over multiple nights are considered to be stationary.) The remaining transient detections form the input to the rest of the pipeline (we do not use MOPS).  We iterate over a set of heliocentric distances, $d\sim25-1500$~au.  For each assumed distance, we carry out a number of steps.  First, we transform the topocentric sky plane positions of transient sources we identify in the \PS\ imaging to those as would be observed from the Sun.   Then, all {\it tracklets} are identified: these are sets of $\geq 2$ detections from the exposures within each individual night that are consistent with linear motion that would be bound to the Sun. No minimum rate of motion is required.
Tracklets with three or more detections are much more likely to be real, as the positions of their constituent detections must be consistent with linear motion at a constant rate.  For every pair of high-confidence tracklets ($\geq 3$ detections) that can be associated with a bound heliocentric orbit, we look for additional supporting tracklets along the great circle defined by those two tracklets.  If an additional tracklet is found, an orbit is fit (using modified routines from the Orbfit package of \citet{Bernstein.2000}) to the set of observations from those three tracklets, and a search is carried out for additional tracklets along the sky plane trajectory defined by that orbit.  As tracklets are found, the orbit is refined and the search continues, recursively.  The algorithm is not greedy; all the different linking possibilities are followed until the set of plausibly connected tracklets is exhausted.  As a final step, single detections that lie within $1\farcs0$ of the sky plane are searched for and incorporated, with astrometric and photometric outliers rejected.

We consider for further investigation any arc of tracklets with detections on at least five separate nights, that has an orbital solution with a reduced chi-squared $\bar{\chi}^2 \lesssim 3$, and that has a range of observation magnitudes that is physically realistic ($\Delta m_w<1.5$). All directions of motion of the resulting orbit are permitted and retained.

\section{Observations}
\label{sec:obs}

The observations of \NAME\ span twelve oppositions. All available photometry is tabulated in Appendix~\ref{sec:appendix} (Table \ref{TAB:DISC}); the astrometry is listed at the Minor Planet Center\footnote{Future citation to the URL of \NAME's discovery MPEC.}.

We initially detected \NAME~in $g$, $r$, and $i$-band observations spanning 2010--2012 from the \PS~$3 \pi$ survey (Table \ref{TAB:DISC}). The absence of $w$-band observations is due to \NAME's $32\arcdeg$ ecliptic latitude at the time of discovery, outside the coverage of the \PS\ $w$-band survey. \NAME~is seen on 12 distinct nights with a total of 24 detections, forming an arc spanning 790 days. \NAME~was retained as a candidate TNO as it passed two tests: a) the residuals to an orbital solution determined with a modified version code of \citet{Bernstein.2000} were consistent with the astrometric uncertainties of the individual detections ($\sim 0\farcs1$), with no outliers; b) the photometric measurements from the \PS~IPP, transformed to $w$-band \citep{Tonry.2012}, spanned only 1.2 magnitudes. 
We visually examined the detections in the \PS\ images for final verification. The photometry in Table~\ref{TAB:DISC} is calibrated to \PS~Data Release 1 \citep[\PS-DR1:][]{PanStarrsDR1:2016} and was measured with the moving-object photometry analysis package TRIPPy \citep{Fraser16a}. 

Although \NAME\ is substantially brighter than the \PS\ detection limits, there is a significant bias against finding such objects with the algorithm we described in \S~\ref{SECN:DET}. 
First, we note that only two tracklets with three or more detections can be seen in Table~\ref{TAB:DISC}.  This is the minimum number for our algorithm.   If either of these tracklets was removed from the data set, the algorithm would not have found \NAME. This could occur moderately often, given the 76\% fill factor of the focal plane~\citep{Chambers.2016}. 
Interestingly, the later of those two tracklets was {\it accidental}: it was found in the overlap between two adjacent fields for which pairs of observations were being taken. The overlap region is 10-20\% of the survey area, depending upon the specific survey pattern.  

We followed up \NAME\ with Sloan $r$-band observations on 2016 July 28--30 with the EFOSC2 camera on the New Technology Telescope \citep[NTT;][]{Buzzoni1984,Snodgrass2008a} atop La Silla, Chile. EFOSC2 uses a LORAL 2048 $\times$ 2048 CCD which was used in 2 $\times$ 2 binning mode. Each binned pixel maps onto an on-sky square 0\arcsec.24 on the side, for a full field of view $4.1\arcmin \times 4.1\arcmin$. The data were subject to standard bias-subtraction and flat-fielding procedures, and the magnitude of \NAME~in each frame was measured using circular aperture photometry and calibrated to PS1-DR1 using tens of field stars. 

\NAME\ is bright enough to be found in archival images with the SSOIS search tool \citep{Gwyn:2012gv}.
We recovered astrometry of \NAME\ from the $griz$ observations of a sequence of 54 s exposures ($riuzg$) made on 2005 May 11. The observations by the 2.5 m telescope of the Sloan Digital Sky Survey (SDSS) in New Mexico were part of SDSS-II \citep{SDSSDR7.2009}. The astrometry of the 0.\arcsec396 pixel images is calibrated to SDSS Data Release 14 \citep{SDSS:2017dr14}. The TNO was close by stars, which prevented photometric measurements.

We also found via SSOIS that \NAME\ was serendipitously imaged in Sloan $g$, $r$ and $z$ on 2014 August 16--18 by the DECalS survey\footnote{\url{http://legacysurvey.org/decamls/}} 
with the Dark Energy Camera (DECam; \citet{DePoy:2008SPIE}) on the 4 m Blanco Telescope in Chile. DECam has 0.\arcsec263 pixels with a 3 deg$^2$ field of view. 
The DECalS astrometry and photometry zeropoints are calibrated\footnote{\url{http://legacysurvey.org/dr3/description/}} to \PS~DR1. 
Photometry, Point-Spread Function (PSF) modelling, and trailed PSF removal were performed with TRIPPy (Table~\ref{TAB:DISC}). No evidence of binarity was found. The five highest signal-to-noise (SNR) and best seeing DECam images are our overall highest-SNR images of \NAME.

\section{Physical properties of \NAME}
\label{SECN:PHYS}

\begin{figure}[htbp]
\centering
\includegraphics[width=\columnwidth]{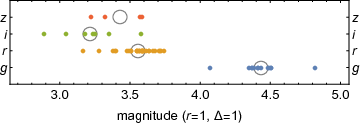}
\caption{%
    \NAME~magnitudes in each band corrected to unit heliocentric and geocentric distance. Gray circles indicate mean magnitudes.
}
\label{FIG:colorS}
\end{figure}

The mean colors of \NAME\ were calculated from the mean magnitudes in each band, after correction to unit heliocentric and geocentric distance (Fig.~\ref{FIG:colorS}). Standard deviations for each band were combined in quadrature to yield the uncertainty. 
Table~\ref{tab:colors} summarizes the derived color measurements. 
The substantial time between observations means rotation could have let potential surface variability intrude \citep{Fraser:2015cx,Peixinho:2015bw}, so we avoid presenting the colors as a coarsely sampled spectrum. 
The $g-r$ and $r-i$ of \NAME~fall along the locus of the known range of TNOs \citep[e.g.][]{Ofek:2012}, classing it as moderately red\footnote{Relative to solar color $g-r = 0.44 \pm 0.02$: \url{sdss.org/dr12/algorithms/ugrizvegasun/}}. 

The visual albedos constrained by thermal measurements for $2 < H_r < 4$ TNOs are wide-ranging, varying from $p =  0.07-0.21$ \citep{Brucker:2009jp,Lellouch:2013cu,Fraser.2014}. \NAME's red $g-r$ color places it among TNOs that have albedos at the higher end of this range \citep{Lacerda:2014wr,Fraser.2014}.
At $H_r=3.44\pm0.10$ (see below), \NAME\ has a diameter of at least 600 km; at the less reflective albedo limit, it could be as large as 900 km.\footnote{We note that the mass of \NAME~is too low to be producing the warp in the mean plane of distant KBOs found by \citet{VolkMalhotra:17}.}
It will thus achieve ellipsoidal hydrostatic equilibrium \citep{Tancredi:2008,Lineweaver:2010te}.

\NAME's $r-z$ color is unusual. To make it consistent with the range of $r-z$ colors exhibited by other TNOs with similar $g-r$ colors as \NAME\ \citep{Pike.2017} requires a $1\sigma$ deviation in both $g-r$ and $r-z$, or a $2 \sigma$ deviation in either. Yet the mean $r-z$ is consistent with the only single-epoch $r-z$ measurement (from DECam; see Table~\ref{tab:colors}).
Alternatively, the negative $r-z$ color is broadly consistent with the presence of methane ice, which exhibits an absorption feature through the $z$-band \citep[see for example][]{Tegler.2007}. For \NAME\ to have a methane-bearing surface would be unusual, as it is expected that objects of this size would rapidly lose that volatile to space on short timescales \citep{Schaller.2007}, in contrast to the larger dwarf planets; e.g. (50000) Quaoar and 2007 OR$_{10}$ are red methane-covered objects \citep{Schaller:2007dta,Brown:2011co}.

\begin{deluxetable}{lLRRR}[htbp]
    \tablecolumns{5}
    \tabletypesize{\small}
    \tablecaption{Optical colors of \NAME 
        \label{tab:colors}
    }
    \tablehead{\colhead{Facility} & MJD & \colhead{$g-r$}  & \colhead{$r-i$} & \colhead{$r-z$} 
    }
    \startdata
    \cutinhead{Mean}
        All          &  &    0.88 \pm 0.21   & 0.34 \pm 0.26  & 0.13 \pm 0.22   \\
    \cutinhead{Single-epoch}
        \PS     & 55711.36    & 1.2 \pm 0.12    &                &                 \\
        DECam   & 56886.00    &                 &                & -0.09 \pm 0.25  \\
        NTT     & 57597.13 & 0.95 \pm 0.13   &                &                 \\
        NTT     & 57599.01 & 0.85 \pm 0.04   &                &                 \\
        NTT     & 57599.04 & 0.74 \pm 0.04   &                &                 \\
        NTT     & 57599.08 & 0.78 \pm 0.03   &                &                 \\
    \enddata
    \tablecomments{For computation of mean color, see text in \S~\ref{SECN:PHYS}.     }
\end{deluxetable}

\begin{figure}[htb]
\centering
\includegraphics[width=0.8\columnwidth]{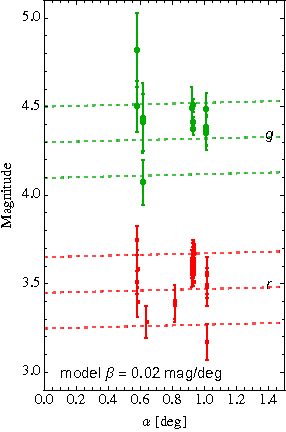}
\caption{%
    Linear phase curve models for \NAME. The slope $\beta=0.02$ mag/deg ($\beta<0.07$ mag/deg, 1-$\sigma$) was fit simultaneously to the $g$ and $r$ photometry (Table~\ref{TAB:DISC}). The dashed lines mark the mean brightness and extent of the variation due to the lightcurve (see Fig.~\ref{FIG:LIGHTCURVE}).
}
\label{FIG:PHASE_CURVE}
\end{figure}

We use the photometry in Table~\ref{TAB:DISC} to constrain the color, phase curve and lightcurve properties for \NAME. The apparent magnitudes were first corrected to unit heliocentric ($r$) and geocentric ($\Delta$) distance by subtraction of $5\log(r\Delta)$. The corrected $g$ and $r$ magnitudes were fit simultaneously with the following model, which accounts for a linear phase darkening and sinusoidal lightcurve variation,
\begin{equation}
H_{g,r}-(g-r)+\beta \alpha + \frac{\Delta m}{2} \sin\left[\frac{2\pi(t-t_0)}{P}\right],
\label{EQN:LC}
\end{equation}
\noindent where $H_{g,r}$ is the absolute $g$ or $r$ magnitude, $\alpha$ is the phase angle, $\beta$ is the linear phase function slope, $\Delta m$, $P$ and $t_0$ are the peak-to-peak variation, period, and offset of the lightcurve, respectively, and $(g-r)$ is a color term subtracted from the $g$ magnitudes only.  As is the case for Earth-based TNO observations, the range of phase angle is limited: $0 < \alpha \lesssim 1\degr$. We found the best-fit $\beta$ to be somewhat dependent on the initialization values for $P$, $\Delta m$, and $\beta$. To solve this, we varied $(g-r)$ linearly in steps of 0.01 mag and at each step generated 200 uniformly distributed random initial values $P\in(16,64)$, $\Delta m\in(0.15,0.60)$, and $\beta\in(0.01,0.50)$. The range of $\beta$ values brackets the slopes seen in TNOs \citep{Rabinowitz:2007AJ....133...26R}. The NTT $r$ band photometry displays a steady brightening by about 0.2 mag over a period of 5 hours, which provides useful limits on the lightcurve variation ($\Delta m>0.2$ mag) and period ($P>10$ hr). We found consistently that shallow phase functions ($\beta<0.05$ mag/deg) fit the data best (see Fig.~\ref{FIG:PHASE_CURVE}). For this reason we decided not to correct the data for phase effects before the lightcurve analysis.

\begin{figure}[htb]
\centering
\includegraphics[width=\columnwidth]{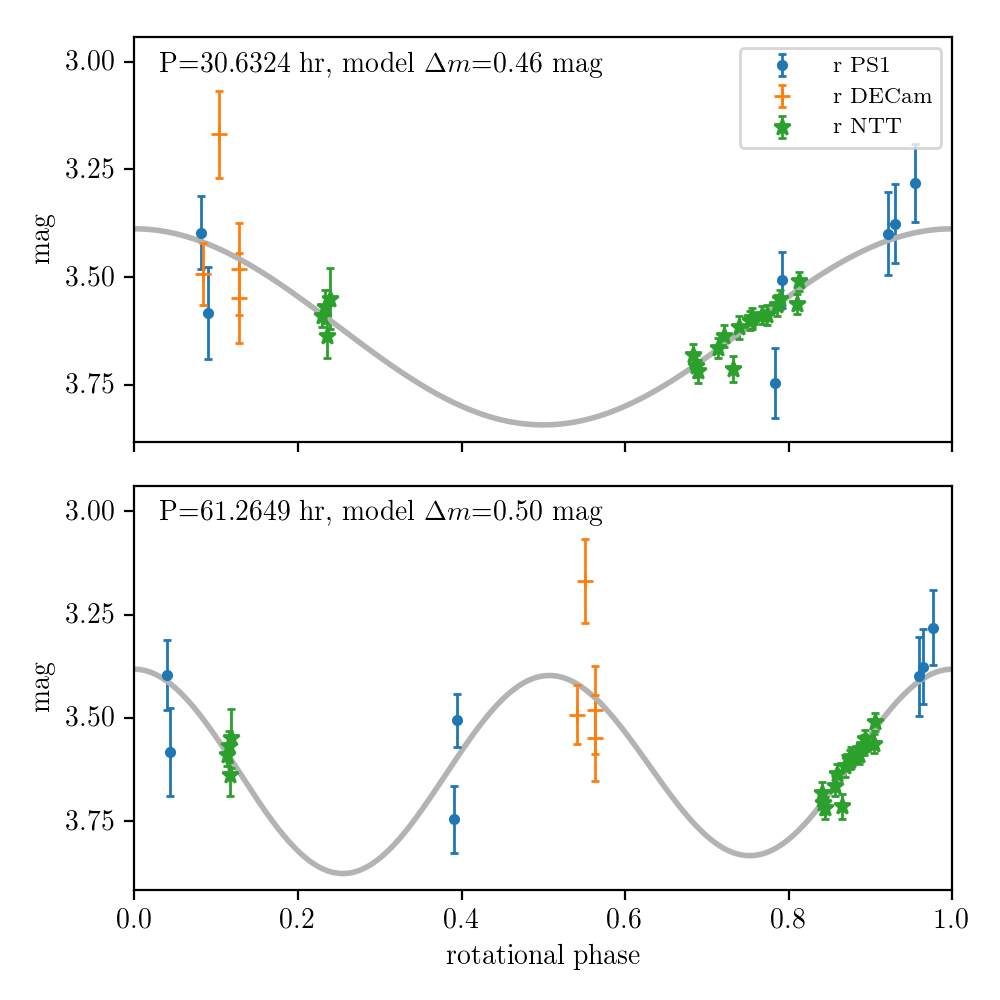}
\caption{%
    Lightcurve of $r$-band photometry of \NAME, color-coded by observing facility (Table~\ref{TAB:DISC}). The single-peaked model implies a nearly spherical shape with a patchy surface spinning with a period $P\approx 32.5$ hr. The corresponding double-peaked solution (period $P\approx 65$ hr) implying an ellipsoidal shape with axes ratio $a/b>1.58$ is less plausible, due to the large material strength required to hold the shape against the crush of gravity.
}
\label{FIG:LIGHTCURVE}
\end{figure}

We employed the Lomb-Scargle algorithm \citep{Press.1989} to measure the lightcurve period using the $r$ band measurements only. The periodogram identified the period $P_1=30.6324$ hr as the best solution, which corresponds to a single-peaked lightcurve, with one maximum and one minimum per rotation. Figure~\ref{FIG:LIGHTCURVE} shows the data phased with $P_1$. The double period $P_2=2P_1=61.2649$ hr is also shown for comparison, in case \NAME~has a symmetric lightcurve.

Both the solutions are consistent with the data and indicate that \NAME~is a slow rotator. The single-peaked solution, $P\sim30.6$ hr, would imply that \NAME~is roughly spherical and has significant albedo patchiness. 
The double-peaked solution would imply an ellipsoidal shape with axes ratio $a/b>1.58$. We find this less plausible, given the large size and slow rotation of \NAME.

Finally, the lightcurve period solutions were used to refine the phase curve, resulting in the linear solution shown in Fig.~\ref{FIG:PHASE_CURVE}, with $\beta=0.02$ mag/deg. We force $\beta>0$ in our fitting algorithm, which sets the lower limit on the phase function slope, and find a 1-$\sigma$ upper limit $\beta<0.07$ deg/mag. We find an absolute $r$ magnitude $H_r=3.44\pm0.10$ and a best-fit color offset $g-r=0.85$, consistent with the $g-r$ values in Table~\ref{tab:colors}.

\section{Orbital Dynamics of \NAME}
\label{SECN:DYN}

We determine the barycentric elements of \NAME\ by fitting observations from six oppositions spanning 2005-2016 (Table~\ref{TAB:FIT}), using the code of \citet{Bernstein.2000}.
The barycentric distance to \NAME\ at discovery in 2010 is $55.019 \pm 0.003$~au.

\begin{deluxetable}{RRRRRR}[htb]
    \tablecolumns{6}
    \tabletypesize{\footnotesize}
    \tablecaption{: Barycentric osculating elements of \NAME\ in the International Celestial Reference System at epoch 2455327.1.\label{TAB:FIT}}
    \tablehead{
    \colhead{$a$ (au)} & \colhead{$e$} & \colhead{$i$ (\degr)} & \colhead{$\Omega$ (\degr)} & \colhead{$\omega$ (\degr)} & \colhead{T$_{\textrm{peri}}$ (JD)} }
    \decimals
    \startdata
        %
        78.307    & 0.49781     &  32.04342     & 147.1722    &  9.824      & 2433937.1 \\
        $\pm$ 0.009 & $\pm$ 0.00005 &  $\pm$ 0.00001  & $\pm$ 0.0002  &  $\pm$ 0.005  &   $\pm$ 0.8 \\
    \enddata
    \tablecomments{Best-fit orbit to the arc of observations 2005--2016 with the method of \citet{Bernstein.2000},  with $1\sigma$ uncertainty estimates from the covariance matrix.
    }
\end{deluxetable}

We generate 100 clones of the orbit of \NAME, 
varying the observations by their respective uncertainties and refitting the orbit, 
hence creating a statistically rigorous bundle of orbits which are all consistent with the observations.
We integrate the best-fit orbit and clones as test-particles in a barycentric system, orbiting in the gravitational field of the Sun and four giant planets. 
We integrate the particles using the adaptive IAS15 integrator \citep{RS15} in the {\sc REBOUND} code of \citet{RL12} for 700 Myr, i.e. around $10^6$ orbits of \NAME.

    \begin{figure}[htb]
    \centering
    \includegraphics[trim = 0mm 0mm 0mm 0mm, clip, angle=0, width=\columnwidth]{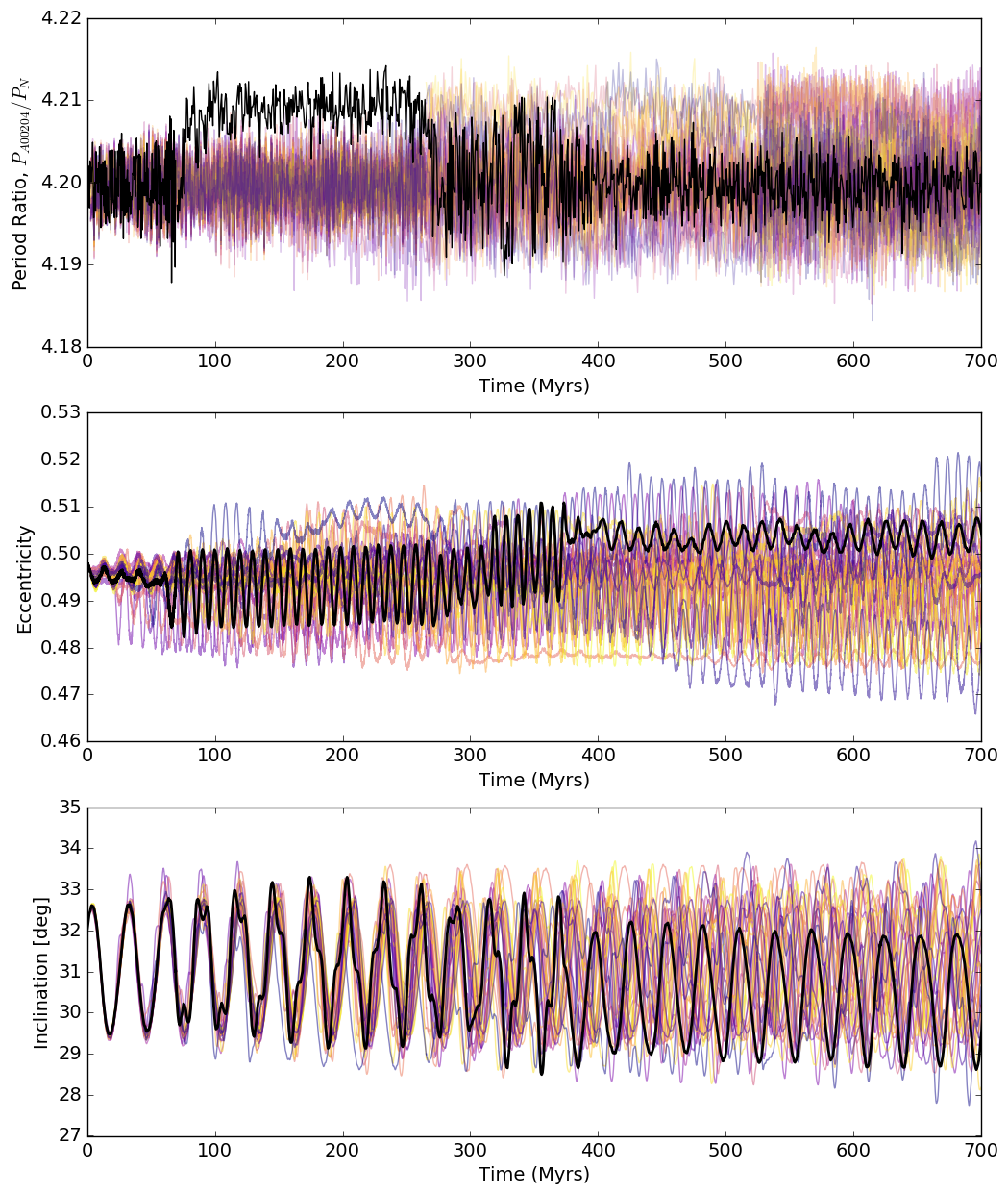}
    \includegraphics[trim = 0mm 0mm 0mm 0mm, clip, angle=0, width=\columnwidth]{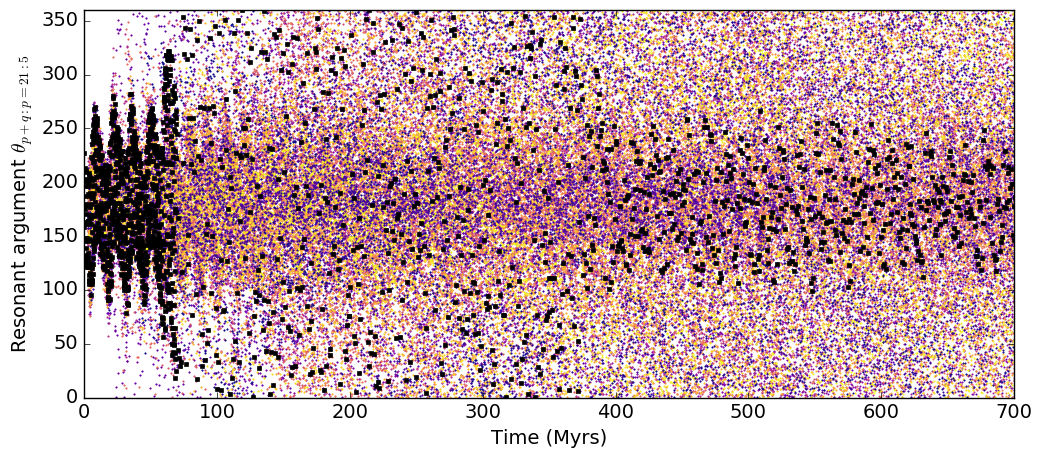}
    \caption{%
    Dynamical evolution of the best-fit orbit for \NAME~(black) and clones (colors) sampled from the covariance matrix of the orbital uncertainties.
    {\bf Top:} Period ratio with Neptune; 
    {\bf 2nd:} Eccentricity; 
    {\bf 3rd:} Inclination; 
    {\bf Bottom: } Resonance angle, $\theta_{p+q:p = 21:5}$ (see Eqn \ref{EQN:RES}).
    The best-fit orbit stably librates in the 21:5 resonance with Neptune for $\sim70$ Myr, before diffusing to a circulating state, and then back to librating. 
    The clones exhibit behavior consistent with that of the  best-fit orbit, switching between resonant and non-resonant configurations on $\sim100$~Myr timescales, with a slow diffusion away from resonance, such that at 700~Myr, $\sim25\%$ remain in resonance.
    }
    \label{FIG:DYNAMICS}
    \end{figure}

The period ratios with Neptune, eccentricity and inclination for the best-fit orbit (black) and a sample of the clones (all other colors) are shown in the top three panels of Figure \ref{FIG:DYNAMICS}.
All of the particles stably orbit with period ratios very close to $4.2$.
We then consider the resonant angle,
\begin{eqnarray}
\theta_{p+q:q} &=& (p+q)\lambda_{out} - p\lambda_{in} - q\varpi_{out},
\label{EQN:RES}
\end{eqnarray}
and plot in the lower panel of Figure \ref{FIG:DYNAMICS} values for Eqn. \ref{EQN:RES} using $p+q:p = 21:5$.
The best-fit orbit exhibits stable libration in the 21:5 resonance for $\sim100$~Myr before diffusing to a circulating configuration with a period ratio $\sim4.21$ with respect to Neptune for $\sim300$~Myr, before returning to a 21:5 resonant configuration for the remainder of the simulation.
We note that $16$ possible resonant angles exist for the 21:5 resonance in which the $-q\varpi_{out}$ term in Eqn.~\ref{EQN:RES} is replaced by $- ( n\varpi_{in} + (q-n)\varpi_{out} )$, with $0\leq n \leq15$. 
We examined all such variations for all best-fit and clone orbits, and found that only the $n=0$ case (i.e. Eqn.~\ref{EQN:RES}) ever exhibits resonance.

The clones of the orbit of \NAME\ display behavior consistent with that of the best-fit orbit.
For the entirety of the simulation, all clones have period ratios with Neptune that remain bounded between $4.185$ and $4.215$ (semi-major axes bounded between $\sim78.1$~au and $\sim78.5$~au) and have pericenters in the range $37.5 - 41.5$~au. 
Most clones move back and forth between resonant and non-resonant configurations, with a slow overall diffusion away from resonance.
\emph{All} clones are resonant at the start of the simulation, and $\sim 6\%$ remain resonant for the entire simulation.
The median time at which clones exit resonance is $\sim110$ Myr.  
At 700~Myr, $\sim25\%$ of clones are resonant.  

The behavior of the best-fit orbit and its clones over the 700~Myr simulation shows that \NAME\ has a metastable orbit, which switches between resonant and non-resonant configurations on hundred-Myr timescales.

\section{Discussion}
\label{SECN:CON}


    \begin{figure*}[htb!]
    \centering
    \includegraphics[trim = 0mm 0mm 0mm 0mm, clip, angle=0,width=0.85\textwidth]{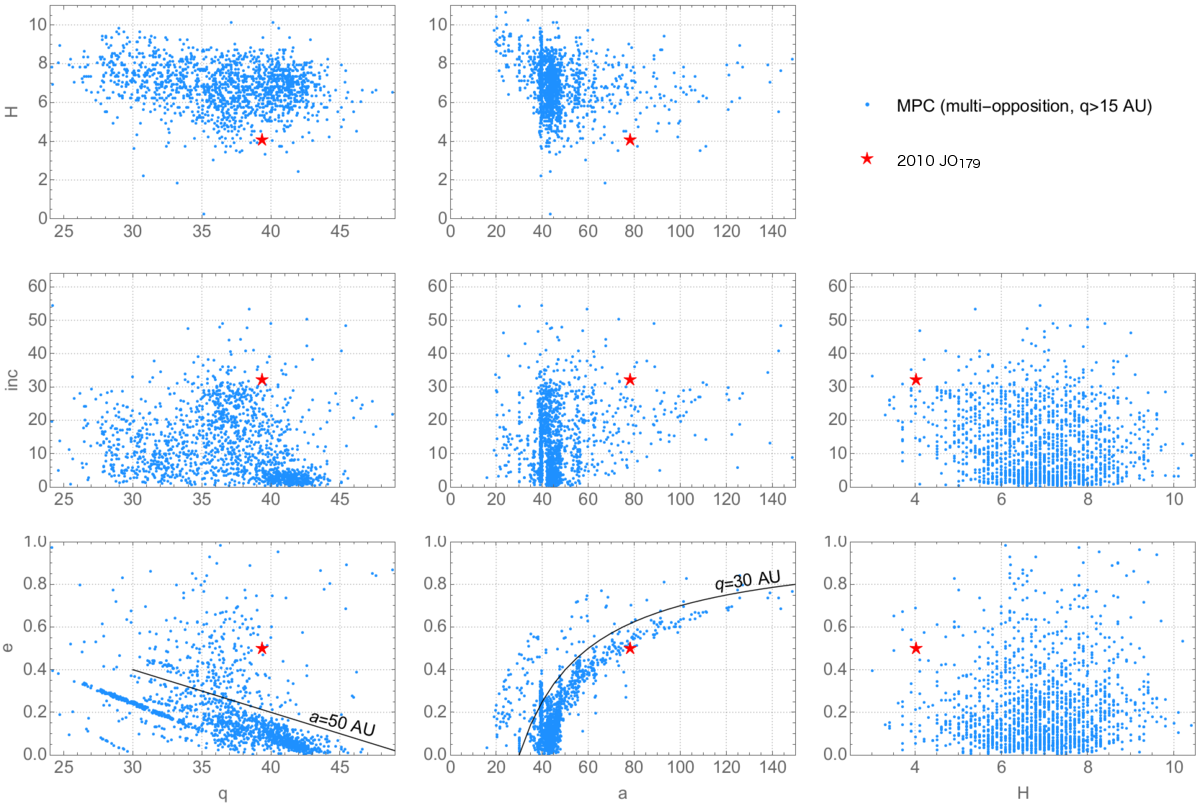}
    \caption{%
    \NAME~placed in context among the other known TNOs.
    }
    \label{FIG:TNO_POPN}
    \end{figure*}

Although \NAME\ is bright for a TNO at $m_r\sim21$, its $32\degr$ orbital inclination and resulting current $\sim 30\degr$ ecliptic latitude is responsible for it not being detected earlier.  It is bright enough to have been detected by the surveys of \citet{Larsen:2007} and \citet{Schwamb.2010}, but fell just outside their sky coverage. It was not detected by \citet{W16}, presumably either because \NAME\ is outside of the region considered in that search, or it was not observed with their required cadence. \NAME\ is thus a good example of how the detection efficiency of a survey is a function of each specific analysis, given a common observational dataset.
Surveys that cover large fractions of the sky, both well away from the ecliptic and to fainter limiting magnitudes, such as are planned for the forthcoming Large Synoptic Survey Telescope \citep{LSST.2009}, will discover many more such objects.  However, archival data will clearly still yield new discoveries with further thorough searching.

For population studies, the detection efficiency of surveys such as \PS\ need to be well characterized, as has been done for the CFEPS and OSSOS surveys \citep{Petit.2011,Bannister:2016a}, to be able to correct for the observational biases. In addition, detailed dynamical classifications of the discoveries, of the type we have done here, need to be made. These permit determination of the intrinsic abundance of TNOs such as \NAME\ \citep[e.g.][]{Pike:2015gn}.  This is the subject of future work with PS1.

With a perihelion distance $q = 39.32$~au (Fig.~\ref{FIG:TNO_POPN}), near the boundary between what are considered to be low perihelion and high perihelion TNOs \citep{Lykawka:2007ff,Gladman.2008}, \NAME\ is a relatively nearby example of what is now being recognized as a very substantial population~\citep{Gomes.2003,Trujillo.2014,Pike:2015gn,KaibSheppard16,Nesvorny16a,Nesvorny16b}. 
In the absence of a survey characterization, we cannot provide an absolute estimate for the 21:5 resonant population, especially given the strong observability bias on the eccentricity distribution of such a large-$a$ population. 
However, we can find a general lower limit. The existence of \NAME\ requires there to be at least one $H_r \simeq 3.5$ TNO. Scaling according to the size distribution of the dynamically excited TNOs \citep{Fraser.2014}, there are at least 6700 objects in the 21:5 that are larger than 100 km in diameter. This would be more numerous than the 3:2 plutinos \citep{Gladman.2012}, and is consistent with the large populations found in other distant resonances \citep{Pike:2015gn,Volk:2016gl}. Aspects of these large populations remain challenging to form in migration scenarios, cf. \citet{Pike:2017nice}.

Neptune's early migration into the outer planetesimal disk emplaced the dynamically excited trans-Neptunian populations, including the resonant and scattering disk \citep{Malhotra:1995dy, Gomes.2003,Gladman.2008}. 
Predicting the details of the emplaced distant populations is an area of active investigation. 
\citet{Nesvorny16b} and \citet{KaibSheppard16}~independently predict that some TNOs with $a> 50$ au and $q > 40$ au should exhibit semi-major axes clustered near and inward, but not within, mean-motion resonances (MMR) with Neptune.
This followed from modelling Neptune's dynamical evolution on an orbit with moderate $e\lesssim0.1$ eccentricity, both under smooth migration and from ``grainy'' gravitational interaction with a small sea of dwarf planets in the initial planetesimal disk \citep{Nesvorny16a}.
In direct contrast, \citet{Pike:2017resonances} predict fairly symmetric trails of high-perihelia populations immediately surrounding distant mean-motion resonances. They assessed emplacement by the Nice model scenario of \citet{Brasser:2013}, with a smoothly migrating, initially high $e=0.3$ Neptune.
While one might hope to constrain the details of migration from the present-day orbit of a distant TNO like \NAME, it unfortunately lies in a region of phase space where its orbit does not distinguish among current model outcomes. Irrespective of whether Neptune's migration was smooth or grainy, fast or slow, the results of \citet{Nesvorny16b}, \citet{KaibSheppard16} and \citet{Pike:2017resonances} all contain TNOs with $a\sim78$~au, $q\sim39$~au, $i\sim30\degr$.

It is also worth considering if \NAME\ could be a more recent arrival to its current orbit.
The $ a > 50 $ au region is filigreed with high-order resonances. These permit ``resonance sticking'', where a TNO's orbit temporarily librates in resonance for tens to hundreds of millions of years, chaotically escapes the resonance, changing in semi-major axis, then sticks, librating within another resonance \citep[e.g][]{Lykawka:2007sticking}. The resonant dynamics during a temporary capture can lead to oscillations in eccentricity and inclination that can then weaken the TNO's interaction with Neptune \citep{Gomes08,ST16a}.
However, \NAME's clones all remain close to its resonance (\S~\ref{SECN:DYN}), implying stability for the age of the Solar System, rather than showing behaviour like that of 2015 RR$_{245}$, which moves between the 9:2 resonance and the scattering disk \citep{Bannister16}. \NAME's clones exhibit similar behaviour to the four known TNOs in and by the 5:1 resonance at $a\sim88$~au \citep{Pike:2015gn}. 
The current orbit of \NAME\ is therefore more likely to be ancient.

\facilities{PS1, NTT (EFOSC2), Sloan, CTIO:Blanco (DECam)}

\software{Astropy, TRIPPy, REBOUND, Matplotlib}

\acknowledgments

MJH and MJP gratefully acknowledge 
NASA Origins of Solar Systems Program grant NNX13A124G 
and grant NNX10AH40G via sub-award agreement 1312645088477, 
BSF Grant 2012384, 
NASA Solar System Observations grant NNX16AD69G, 
as well as support from the Smithsonian 2015 CGPS/Pell Grant program.
MTB appreciates support from UK STFC grant ST/L000709/1.

The Pan-STARRS1 Surveys (PS1) have been made possible through contributions of the Institute for Astronomy, the University of Hawaii, the Pan-STARRS Project Office, the Max-Planck Society and its participating institutes, the Max Planck Institute for Astronomy, Heidelberg and the Max Planck Institute for Extraterrestrial Physics, Garching, The Johns Hopkins University, Durham University, the University of Edinburgh, Queen's University Belfast, the Harvard-Smithsonian Center for Astrophysics, the Las Cumbres Observatory Global Telescope Network Incorporated, the National Central University of Taiwan, the Space Telescope Science Institute, the National Aeronautics and Space Administration under Grant No. NNX08AR22G issued through the Planetary Science Division of the NASA Science Mission Directorate, the National Science Foundation under Grant No. AST-1238877, the University of Maryland, and Eotvos Lorand University (ELTE).

This work is based in part on observations collected at the European Organisation for Astronomical Research in the Southern Hemisphere under ESO programmes 194.C-0207(H), and on observations at Cerro Tololo Inter-American Observatory, National Optical Astronomy Observatory, which is operated by the Association of Universities for Research in Astronomy (AURA) under a cooperative agreement with the National Science Foundation. 

The computations in this paper were run on the Odyssey cluster supported by the FAS Science Division Research Computing Group at Harvard University.

This research used the facilities of the Canadian Astronomy Data Centre operated by the National Research Council of Canada with the support of the Canadian Space Agency.


\appendix
\section{Observational Data}
\label{sec:appendix}

\startlongtable
\begin{deluxetable*}{lDcRRc}
    \tabletypesize{\small}
    \tablecolumns{6}
    \tablecaption{Photometry of \NAME \label{TAB:DISC}}
    \tablehead{\colhead{Date (UTC)} & \twocolhead{Date (MJD)} & \colhead{Filter} & \colhead{$m_{filter}$} & \colhead{Exp (s)} & \colhead{Facility}}
    \decimals
    \startdata
    2010-05-10 13:51:55.9 & 55326.577730 & i & 20.62 \pm 0.10 & 45 & PS1 \\ 
    2010-05-10 14:08:27.7 & 55326.589210 & i & 20.26 \pm 0.10 & 45 & PS1 \\ 
    2010-05-11 13:50:39.0 & 55327.576840 & r & 20.77 \pm 0.09 & 40 & PS1 \\ 
    2010-05-11 14:07:02.2 & 55327.588220 & r & 20.95 \pm 0.11 & 40 & PS1 \\ 
    2010-06-04 08:19:36.5 & 55351.346950 & i & 20.95 \pm 0.09 & 45 & PS1 \\ 
    2010-06-04 08:21:36.6 & 55351.348340 & i & 20.61 \pm 0.07 & 45 & PS1 \\ 
    2010-06-04 08:37:05.4 & 55351.359090 & i & 20.73 \pm 0.08 & 45 & PS1 \\ 
    2010-06-04 08:39:04.6 & 55351.360470 & i & 20.55 \pm 0.07 & 45 & PS1 \\ 
    2010-06-08 11:51:56.2 & 55355.494400 & r & 20.66 \pm 0.09 & 40 & PS1 \\ 
    2010-07-01 10:15:11.8 & 55378.427220 & r & 20.78 \pm 0.10 & 40 & PS1 \\ 
    2010-07-01 10:31:53.2 & 55378.438810 & r & 20.76 \pm 0.09 & 40 & PS1 \\ 
    2011-05-30 08:33:38.9 & 55711.356700 & g & 22.22 \pm 0.21 & 43 & PS1 \\ 
    2011-05-30 08:48:54.7 & 55711.367300 & g & 21.91 \pm 0.15 & 43 & PS1 \\ 
    2011-05-30 09:05:46.5 & 55711.379010 & r & 21.15 \pm 0.08 & 40 & PS1 \\ 
    2011-05-30 09:20:08.7 & 55711.388990 & r & 20.91 \pm 0.07 & 40 & PS1 \\ 
    2011-06-07 12:20:36.4 & 55719.514310 & g & 21.48 \pm 0.13 & 43 & PS1 \\ 
    2011-06-07 12:35:18.5 & 55719.524520 & g & 21.82 \pm 0.16 & 43 & PS1 \\ 
    2011-08-15 08:10:48.6 & 55788.340840 & i & 20.48 \pm 0.21 & 45 & PS1 \\ 
    2012-06-08 08:00:09.2 & 56086.333440 & g & 21.87 \pm 0.20 & 43 & PS1 \\ 
    2014-08-16 00:22:58.0 & 56885.015949 & g & 22.02 \pm 0.09 & 84 & DECam \\ 
    2014-08-16 01:07:29.5 & 56885.046869 & g & 21.90 \pm 0.09 & 88 & DECam \\ 
    2014-08-16 01:37:17.4 & 56885.067563 & g & 21.92 \pm 0.10 & 82 & DECam \\ 
    2014-08-16 01:48:27.6 & 56885.075320 & g & 21.88 \pm 0.09 & 87 & DECam \\ 
    2014-08-16 23:11:35.6 & 56885.966385 & z & 21.12 \pm 0.24 & 122 & DECam \\ 
    2014-08-17 00:09:24.4 & 56886.006532 & r & 21.03 \pm 0.07 & 70 & DECam \\ 
    2014-08-17 00:43:23.8 & 56886.030136 & r & 20.70 \pm 0.10 & 72 & DECam \\ 
    2014-08-17 01:28:22.7 & 56886.061374 & r & 21.02 \pm 0.11 & 69 & DECam \\ 
    2014-08-17 01:30:00.8 & 56886.062509 & r & 21.08 \pm 0.10 & 71 & DECam \\ 
    2014-08-18 00:29:15.7 & 56887.020321 & z & 20.86 \pm 0.09 & 121 & DECam \\ 
    2014-08-18 01:29:05.1 & 56887.061864 & z & 20.76 \pm 0.10 & 124 & DECam \\ 
    2014-08-18 01:31:38.7 & 56887.063642 & z & 21.11 \pm 0.13 & 127 & DECam \\ 
    2016-07-28 02:51:47.5 & 57597.119300 & r & 21.18 \pm 0.03 & 300 & NTT \\ 
    2016-07-28 02:57:24.5 & 57597.123200 & r & 21.15 \pm 0.04 & 300 & NTT \\ 
    2016-07-28 03:03:01.4 & 57597.127100 & r & 21.22 \pm 0.05 & 300 & NTT \\ 
    2016-07-28 03:08:55.7 & 57597.131200 & r & 21.13 \pm 0.07 & 300 & NTT \\ 
    2016-07-28 03:14:32.6 & 57597.135100 & g & 22.08 \pm 0.12 & 300 & NTT \\ 
    2016-07-29 23:24:08.6 & 57598.975100 & r & 21.27 \pm 0.03 & 300 & NTT \\ 
    2016-07-29 23:29:37.0 & 57598.978900 & r & 21.29 \pm 0.03 & 300 & NTT \\ 
    2016-07-29 23:35:13.9 & 57598.982800 & r & 21.30 \pm 0.03 & 300 & NTT \\ 
    2016-07-30 00:20:26.9 & 57599.014200 & r & 21.2 5 \pm 0.02 & 300 & NTT \\ 
    2016-07-30 00:26:12.5 & 57599.018200 & g & 22.09 \pm 0.03 & 300 & NTT \\ 
    2016-07-30 00:31:58.1 & 57599.022200 & r & 21.22 \pm 0.03 & 300 & NTT \\ 
    2016-07-30 00:54:08.6 & 57599.037600 & r & 21.30 \pm 0.03 & 300 & NTT \\ 
    2016-07-30 00:59:54.2 & 57599.041600 & g & 21.99 \pm 0.04 & 300 & NTT \\ 
    2016-07-30 01:05:31.2 & 57599.045500 & r & 21.20 \pm 0.03 & 300 & NTT \\ 
    2016-07-30 01:31:35.0 & 57599.063600 & r & 21.19 \pm 0.02 & 300 & NTT \\ 
    2016-07-30 01:36:11.5 & 57599.066800 & r & 21.18 \pm 0.03 & 300 & NTT \\ 
    2016-07-30 01:58:39.4 & 57599.082400 & r & 21.17 \pm 0.02 & 300 & NTT \\ 
    2016-07-30 02:04:25.0 & 57599.086400 & g & 21.95 \pm 0.03 & 300 & NTT \\ 
    2016-07-30 02:10:10.6 & 57599.090400 & r & 21.17 \pm 0.02 & 300 & NTT \\ 
    2016-07-30 02:33:04.3 & 57599.106300 & r & 21.15 \pm 0.03 & 300 & NTT \\ 
    2016-07-30 02:37:40.8 & 57599.109500 & r & 21.14 \pm 0.02 & 300 & NTT \\ 
    2016-07-30 03:16:59.5 & 57599.136800 & r & 21.15 \pm 0.02 & 300 & NTT \\ 
    2016-07-30 03:21:36.0 & 57599.140000 & r & 21.09 \pm 0.02 & 300 & NTT \\ 
    \enddata
    \tablecomments{\PS\ and DECam photometry measured with TRIPPy \citep{Fraser16a}, NTT photometry measured with circular aperture photometry.}
\end{deluxetable*}




\end{document}